# Coexistence of near-$E_F$ flat band and van Hove singularity in a two-phase superconductor

Xuezhi Chen[1,2,3] [*], Le Wang[4,5] [*], Jun Ishizuka[6] [*], Kosuke Nogaki[7], Yiwei Cheng[1,2,3], Fazhi Yang[2], Renjie Zhang[2,3,8], Zhenhua Chen[9], Fangyuan Zhu[9], Youichi Yanase[7] [#], Baiqing Lv[2,10,11] [#], Yaobo Huang[1,3,9] [#]

[1] *Shanghai Institute of Applied Physics, Chinese Academy of Sciences, Shanghai 201800, China*
[2] *Tsung-Dao Lee Institute, Shanghai Jiao Tong University, Shanghai 200240, China*
[3] *University of Chinese Academy of Sciences, Beijing 100049, China*
[4] *Shenzhen Institute for Quantum Science and Engineering, Southern University of Science and Technology, Shenzhen 518055, China*
[5] *International Quantum Academy, Shenzhen 518048, China*
[6] *Faculty of Engineering, Niigata University, Ikarashi, Niigata 950-2181, Japan*
[7] *Department of Physics, Kyoto University, Kyoto 606-8502, Japan*
[8] *Beijing National Laboratory for Condensed Matter Physics and Institute of Physics, Chinese Academy of Sciences, Beijing, 100190, China*
[9] *Shanghai Synchrotron Radiation Facility, Shanghai Advanced Research Institute, Chinese Academy of Sciences, 201204 Shanghai, China*
[10] *School of Physics and Astronomy, Shanghai Jiao Tong University, Shanghai 200240, China.*
[11] *Zhangjiang Institute for Advanced Study, Shanghai Jiao Tong University, Shanghai 200240, China.*

[*] These authors contributed equally to this work.
[#] Corresponding authors: yanase@scphys.kyoto-u.ac.jp, baiqing@sjtu.edu.cn, huangyaobo@sari.ac.cn

**Abstract**

In quantum many-body systems, particularly, the ones with large near-$E_F$ density states, like flat bands or van Hove singularity (VHS), electron correlations often give rise to rich phase diagrams with multiple coexisting/competing orders occurring at similar energy scales. The recently discovered locally noncentrosymmetric heavy fermion superconductor $CeRh_2As_2$ has stimulated extensive attention due to its unusual H-T phase diagram, consisting of two-phase superconductivity, antiferromagnetic order, and possible quadrupole-density wave orders. However, despite its great importance, the near-$E_F$ electronic structure remains experimentally elusive. Here, we provide this key information by combining soft X-ray and vacuum ultraviolet (VUV) angle-resolved photoemission spectroscopy measurements and atom-resolved DFT+U calculations. With bulk-sensitive soft X-rays, we reveal quasi-2D hole- and 3D electron- pockets with a pronounced nesting feature. On the other hand, under VUV light, the Ce Kondo coherence peaks are resolved in a wide photon energy range (90 eV - 205 eV), deviating from the typical resonance behavior of heavy fermion materials. Most importantly, we observe a symmetry-protected fourfold VHS coexisting with the Ce $4f^1_{5/2}$ flat bands near the $E_F$, which, to the best of our knowledge, has never been reported before. Such a rare coexistence is expected to lead to a large density of states at the zone edge, enhancement in electron correlations, and a large upper critical field of the odd-parity superconducting phase. Uniquely, it will also result in a new type of *f*-VHS hybridization that alters the order and fine electronic structure of the symmetry-protected VHS and flat bands. These peculiarities offer important dimensions for understanding the reported rich phase diagram and are discussed as an origin of superconductivity with two phases. Our findings not only provide key insights into the nature of multiple phases in $CeRh_2As_2$, but also open up new prospects for exploring the novelties of many-body systems with *f*-VHS hybridization.

## I. INTRODUCTION

In correlated electron systems, due to the interplay of charge, lattice, spin, and orbital degrees of freedom, multiple orders exist with close energy, temperature, or time scales. Understanding and engineering these different orders has always been the major theme of condensed matter physics. One important aspect is the electronic structure near the Fermi energy $E_F$. It is generally believed that a large density of states (DOS), often resulting from the flat band or van Hove singularity (VHS), can enhance the many-body interactions and result in exotic correlated phenomena. The two most well-known and extensively studied examples are cuprates and heavy fermion superconductors [1–14]. In both systems, the presence of VHS or flat band plays an important role in driving unconventional superconductivity as well as controlling the interplay among superconductivity, spin and charge density waves, or nematic and orbital orders.

Recently, the locally noncentrosymmetric heavy fermion superconductor, CeRh$_2$As$_2$ has attracted much attention [15–37]. Apart from the Kondo $f$-$d$ electron band hybridization, it exhibits the following peculiarities. 1) The most striking feature is the c-axis field-induced two-phase superconductivity SC1 and SC2, as illustrated in Fig. 1 [15,34]. As we know, multiphase unconventional superconductors are rare in nature. CeRh$_2$As$_2$ stands out not only because of the existence of the SC1 and SC2 phases but also due to the odd-parity nature of SC2, providing a feasible platform for realizing topological crystalline superconductors and Majorana fermions [28,38]. It's also worth highlighting that the corresponding upper critical field of SC2 is up to 14 T, well exceeding the Pauli-limiting field of BCS superconductors. 2) Slightly above $T_c$, another non-magnetic order occurs and persists into the superconducting state. Such coexisting order was interpreted as a quadrupole-density wave (QDW) [16]. 3) Surprisingly, recent nuclear quadrupole resonance (NQR) experiments revealed an antiferromagnetic (AFM) order with an odd-parity multipole inside the SC phase [17,19], beyond our conventional wisdom of superconductivity and magnetism, where the coexistence has only been observed when $T_N > T_c$ [39,40]. Finally, all these coexisting phases emerge from a non-Fermi-liquid state below 4 K, indicating the proximity to a quantum critical point [15].

The above results demonstrate a rare coexistence of $f$-electron flat band, two-phase superconductivity, odd-parity magnetic multipole, and putative QDW in CeRh$_2$As$_2$. Theoretically, the unique crystal and near $E_F$ electronic structures are believed to be the main driving force [23,26,27,30–32,41–43] of these phenomena. As illustrated in Fig. 1, CeRh$_2$As$_2$ crystallizes in the CaBe$_2$Ge$_2$-type tetragonal structure with space group P4/nmm [44]. It consists of vertical stacking 2D Ce layers and Rh$_2$As$_2$ blocks. Notably, the Ce layer is located between two inequivalent Rh$_2$As$_2$ blocks (i.e., Rh-As-Rh and As-Rh-As). Therefore, the inversion symmetry is locally broken at the Ce sites, while the global inversion symmetry is still maintained with the inversion center in the middle of the two Ce atoms. The locally noncentrosymmetric crystal structure and the consequent Rashba spin-orbit coupling (SOC), which is larger than the interlayer hopping as a result of large near-edge DOS, are crucial for the AFM order and the field-induced parity transition [23,26,27,29,31,32]. On the other hand, the pronounced nesting features of the Fermi surfaces play an essential role in stabilizing QDW order [16]. It has been increasingly clear that the near-$E_F$ electronic structure is the key ingredient for clarifying the rich phases in CeRh$_2$As$_2$. However, the momentum-resolved band structure so far remains elusive experimentally. In particular, neither angle-resolved photoemission spectroscopy (ARPES), scanning tunneling microscopy (STM), nor quantum oscillation measurement has been reported.

In this work, we present the first report on the band structure from a combination of bulk-sensitive soft X-ray and resonant vacuum ultraviolet (VUV) ARPES. The observation of quasi-2D hole- and 3D electron-pockets with similar sizes favors a nesting scenario. Uniquely, we unravel a VHS at the Brillouin zone (BZ) edge X point, hybridizing with the 4 $f_{5/2}^1$ bands in the vicinity of the $E_F$. Our study raises new possibilities of interplay between large-density flat bands and VHS, offering an important dimension for how these hybridizing electrons entangle in various symmetry-broken phases.

## II. RESULTS

We first investigate the bulk electronic structure by performing systematic soft x-ray ARPES measurements on the (001) surface. The employment of soft X-rays improves both the bulk sensitivity and the intrinsic $k_z$ resolution, allowing the accurate navigation of the bulk bands in the 3D BZ by photon-energy-dependent measurements. Indeed, the measured constant energy map in the vertical Γ-M-Z-A plane exhibits a weak modulation along the $k_z$ direction with a period of $2\pi/c$ [Fig. 2(a)], where $c$ is the lattice constant, confirming their bulk nature. Figs. 2(b), 2(c) displays the measured Fermi surface and band dispersion within the $k_z = 0$ plane. One can identify two ring-like hole pockets centered at the Γ point shown curvature plots along Γ-M, labeled as $\alpha$, and $\beta$ [Fig. 2(d)]. Correspondingly, Figs. 2(e), 2(f) also exhibits the two hole bands along the Z-R direction, corroborating with Fig. 2(a), and demonstrates the quasi-2D nature of $\alpha$ and $\beta$. In addition to the hole pockets, Figs. 2(e), 2(f) also shows one electron pocket surrounding the Z point, labeled as $\gamma'$ [Fig. 2(f)]. The absence of an electron pocket in Fig. 2(c) suggests that it has a strong dispersion along the $k_z$ axis. Fig. 2(h) summarizes all the resolved bands, and it is clear that $\gamma'$ matches the outer hole pockets $\beta$ well. As proposed by Ref. [16], such a pronounced nesting feature can account for the QDW. Note that the hybridization of multiple bands near $E_F$ is expected to result in intricate Fermi surfaces, beyond the resolving power of current soft X-ray ARPES experiments.

Moving on the Fermi surface near the X point, we display the high-resolution ARPES intensity plots in Figs. 3(a), 3(b) and corresponding curvature plots in Figs. 3(d), 3(e) along the two perpendicular directions passing through X. Whereas Figs. 3(a), (b) shows an electron bottom centered at the X point, a hole-like band is observed along the X-M direction. Consistently, the electron bottom shifts down instead of up as one moves the cut off the high symmetry line Γ-X direction [Fig. 3(g)]. These observations directly demonstrate the existence of a VHS at the X point, as illustrated in Fig.

3(g). The VHS, locating only approximately 30 meV below the $E_F$, will surely contribute to the exotic phenomena in CeRh$_2$As$_2$.

To further understand the near-$E_F$ states and the VHS, we performed systematic orbital- and atom-resolved DFT+U calculations. The DFT results, in good agreement with the previous reports, are summarized in Fig. 3(h). Comparing the calculated band structure with the experimental data, we came to the following conclusions. First of all, Fig. 3(h) shows hole bands of α, β, α', and β' and an electron band of γ' near $E_F$, affirming our results in Fig. 2. Secondly, one can identify several VHSs at the X point, and these VHSs are actually fourfold Dirac nodal points protected by the nonsymmorphic glide symmetry [45]. In other words, the VHSs are robust against crystal-symmetry preserved perturbations. Thirdly, the SOC-induced band splitting is best seen at the BZ edge, such as in the X-M and R-A directions in Fig. 3(h). While our ARPES measurements cannot access these above $E_F$ bands, we resolved two hole bands that are dominated by the two non-equivalent Rh atoms, as highlighted by the black arrows in Fig. 3(f), further confirming the lack of local inversion symmetry at the Ce sites. Last but not least, although the calculated electronic structure, especially the one with U = 2 eV, is generally consistent with the ARPES results, there are still some notable discrepancies. In particular, the measured VHS is much closer to the $E_F$ compared to Fig. 3(h), suggesting the additional band renormalization due to interactions.

One hallmark of heavy fermion materials is Kondo hybridization at low temperatures. Specifically, in CeRh$_2$As$_2$, the Kondo coherence has been hinted by a broad maximum in resistivity at a temperature ~ 40 K [15], however, direct spectroscopic evidence has been lacking. To visualize the $c$-$f$ electron hybridization, we conducted detailed temperature- and photon energy-dependent VUV ARPES measurements. The spectra taken at 20 K and near the 4$d$ - 4$f$ resonance photon energy (~ 121 eV) show two flat features around binding energy of − 0.25 eV and near $E_F$, providing direct evidence for the existence of $4f_{7/2}^1$ and $4f_{5/2}^1$ Kondo resonance peaks, as expected for Ce-based Kondo systems [11–14]. Unexpectedly, the two flat bands are visible in a wide photon energy range ranging from 90 eV to more than 200 eV in Fig. 4(c), in sharp contrast with other heavy fermion materials, where coherence peaks are only visible near the resonance energy. More theoretical efforts are needed to understand such anomalous behavior. One naive picture could be because these 2D $f$-electron bands are more itinerant in CeRh$_2$As$_2$ [11], therefore they can be seen at off-resonant energy, say, 115 eV (see more details in Fig. 5 in the Appendix A). Moving to the Kondo coherence, the measured flat bands exhibit clear enhancement of spectral intensity at the crossing points of the conduction bands (highlighted by red arrows in Fig. 4(d)), signifying the $c$-$f$ hybridization. The hybridization is better illustrated by the coherence peak in the energy distribution curves (EDCs), as shown in Fig. 4(f). Interestingly, our temperature-dependent measurements show that the coherence peak persists up to more than 120 K [Fig. 6 in the Appendix B]. This indicates that the localized-to-itinerant transition happens at a much higher temperature in CeRh$_2$As$_2$ like CeCoIn$_5$ [13], compared to its Kondo temperature of 40 K.

# III. CONCLUSION AND DISCUSSION

Taken together, our soft X-ray and VUV ARPES results show the rare coexistence of symmetry-protected fourfold VHS and $f$-electron flat bands in $CeRh_2As_2$. To the best of our knowledge, such coexisting has not been reported, and $CeRh_2As_2$ thus provides a unique platform to study the relationship between the flat band, symmetry-protected VHS, and multiple broken-symmetry states. Particularly, our observation provides key information for elucidating the long-sought low-energy excitations and the novel two-phase superconductivity. First of all, generally speaking, the flat band VHS gives rise to a large DOS, which implies that more electrons contribute to the low-energy excitations. As a consequence, many-body interactions are enhanced, leading to a multitude of competing/coexisting phases, as evidenced by cuprates, heavy fermion, and the recently discovered moiré superconductors [46-48]. In this regard, it's reasonable to ascribe the two superconducting phases in $CeRh_2As_2$ to the coexistence of flat band and VHS in a locally noncentrosymmetric crystal structure. Indeed, the absence of unconventional SC in $LaRh_2As_2$ confirms the crucial role of $f$-electron correlation in driving the unconventional SC [49]. Besides the enhanced electron correlations, it has been recognized that a large orbital limit of the critical field $H_{c2}^{orb}$ is essential for the observed two-phase superconductivity in $CeRh_2As_2$ [50]. Theoretically, $H_{c2}^{orb} = \phi_0/(2\pi\xi^2)$, $\xi = v_F/Tc$. Therefore, small $v_F$ of both the flat band and VHS naturally leads to large $H_{c2}^{orb}$. Moreover, the VHS-induced large DOS at the BZ edge is also believed to enhance the field-induced even- to odd-parity transition [23]. The above three clues demonstrate that the coexistence of flat band and fourfold VHS is responsible for the observed large critical field and the even- to odd-parity transition in $CeRh_2As_2$. Apart from stabilizing an odd-parity superconducting state, the presence of VHS leads to divergences of the (spin and/or charge) susceptibilities, which might be related to the AFM and QDW order [10,16,17]. In addition, the application of magnetic field, carrier doping, or pressure, can lead to a Lifshitz transition at the X point, which can change the topological character of $CeRh_2As_2$ [28,30].

# IV. OUTLOOK

Looking forward, our findings also open a new avenue for investigating the interplay between large-DOS flat band and symmetry-protected fourfold VHS, both theoretically and experimentally. As illustrated in Fig. 4(h) compared to conventional $c$-$f$ hybridization in Fig. 1, which is characterized by a gap opening at the crossing points, $f$-VHS hybridization has much richer consequences: (i) it leads to the transition to a higher-order VHS [51]; (ii) it also alters the low-energy excitation near the symmetry-protected Dirac nodal line; (iii) it shifts the energy location of the symmetry-protected VHS. Beyond the $f$-VHS hybridization, it has been shown that spin-triplet pairing interaction is induced by quantum geometry, which is enhanced in topological bands [52]. Thus, the Dirac nodal line might produce spin-triplet Cooper pairs. When spin-triplet Cooper pairs are admixed with spin-singlet ones due to locally noncentrosymmetric crystal structure, the upper critical field is enhanced [53], consistent with the experimental observation in CeRh$_2$As$_2$ [15]. All these characteristics await experimental verifications. In particular, Sub-kelvin and ultra-high resolution ARPES and STM measurements are ideal tools for resolving the fine features of *f*-VHS hybridization, topological transition, as well as their interaction with the observed multiple phases in CeRh$_2$As$_2$.

## V. METHODS

### A. SAMPLE SYNTHESIS

Single crystals of CeRh$_2$As$_2$ were grown by Bi-flux method. The starting materials Cerium, Rhodium, Arsenic and Bismuth were placed in an alumina crucible with a molar ratio of 1.5:2:2.1:30, and subsequently sealed into a fully evacuated quartz tube. The crucible was heated to 1130 °C for 10 hours, and then slowly cooled down to 700 °C for 200 hours. Black single crystals were gained by spinning off the Bi flux in a centrifuge.

### B. ANGLE_RESOLVED PHOTOEMISSION SPECTROSCOPY

High-resolution ARPES measurements were performed at the "Dreamline" beamline of the Shanghai Synchrotron Radiation Facility (SSRF) with a Scienta Omicron DA30L analyzer. The photon energy ranged from 90 eV to 510 eV, and the combined (beamline and analyzer) experimental energy resolution was 15 to 60 meV. The angular resolution of the DA30L analyzer was 0.1°. The beam spot had an approximate cross-sectional size of 100 μm by 100 μm. The chemical potential was determined from the spectra of a polycrystalline gold. Fresh surfaces were obtained by cleaving CeRh$_2$As$_2$ crystals *in-situ* in an ultrahigh vacuum base pressure was maintained below $5.6\times10^{-11}$ Torr and the pressure was below $1\times10^{-10}$ Torr during the temperature dependence ARPES experiments.

### C. DETAILS OF DFT CALCULATION

We perform the DFT calculations for paramagnetic CeRh$_2$As$_2$ using the WIEN2k package [55]. We use the full potential linearized augmented plane wave + local orbitals method within the generalized gradient approximation. The Brillouin zone sampling is performed with 16×16×7 points. We choose the muffin-tin radii (RMT) of 2.5, 2.34, and 2.23 a.u. for Ce, Rh, and As, respectively. The maximum reciprocal lattice vector $K_{max}$ is given by $R_{MT}K_{max} = 8$. The spin-orbit coupling is included in all calculations. For DFT+U, we set Hund's coupling $J = 0$ and subtract the double-counting correlation by the around mean-field formula [56].

## ACKNOWLEDGEMENTS

We thank Gexing Qu and Bingjie Chen for the assistance in the ARPES experiments. We thank Hong Ding, Yifeng Yang, Noah F. Q. Yuan and Xiaoyan Xu for fruitful discussions. Y.B.H. acknowledges support by the CAS Pioneer "Hundred Talents Program" (type C). B.Q.L. acknowledges from the TDLI starting up grant, the National Natural Science Foundation of China, Shanghai Natural Science Fund for Original Exploration Program (23ZR1479900), and Shanghai talent Program. L. W. acknowledges support by the National Natural Science Foundation of China (Grant No. 12204223). Y.Y. acknowledges support by JSPS KAKENHI (Grant Nos. JP22H01181, JP22H04933). K.N. was supported by JSPS KAKENHI (grant no. JP21J23007).

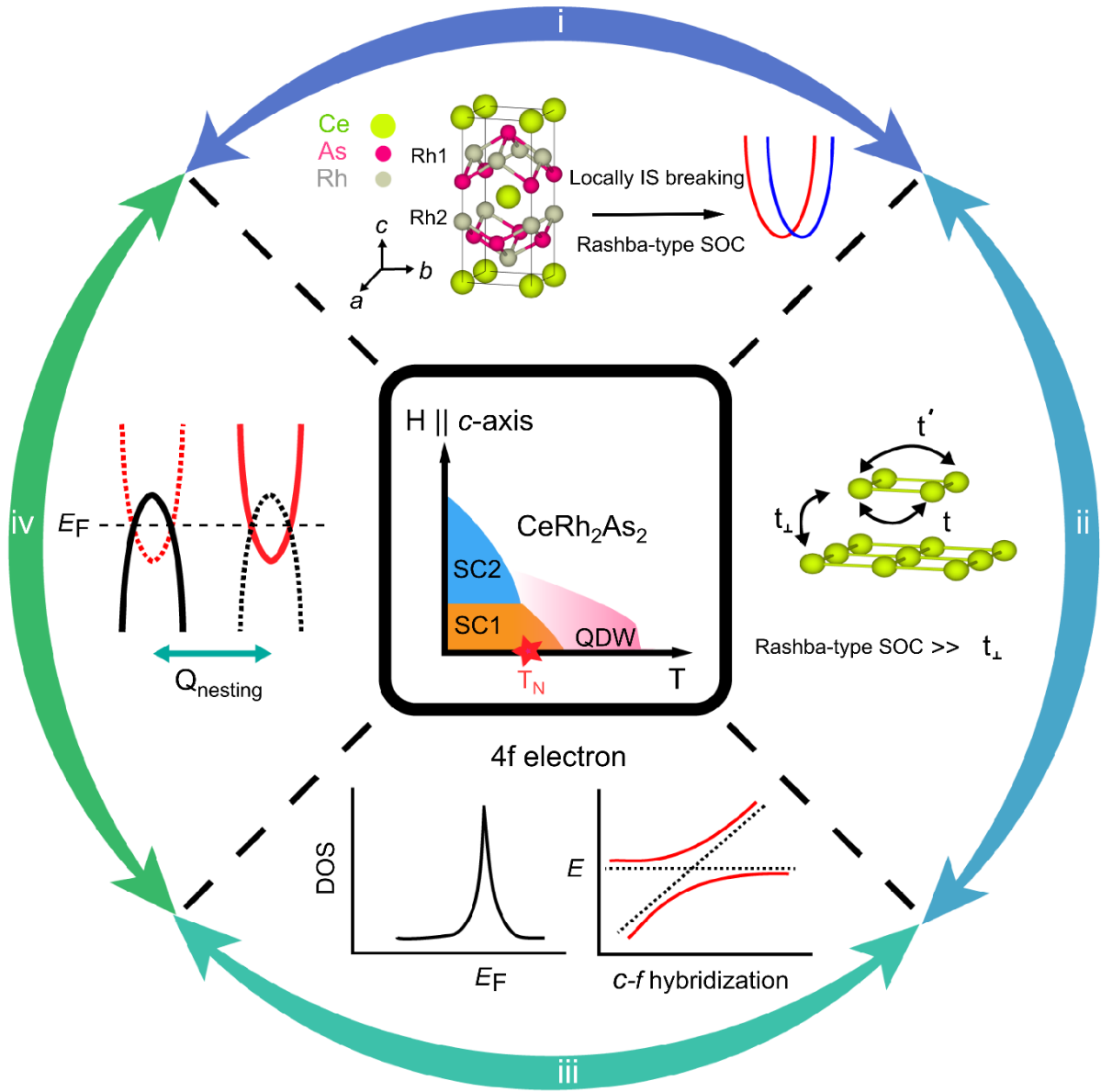

Fig. 1. Schematic H-T phase diagram of CeRh$_2$As$_2$. The phase transition between two superconducting phases has been observed with H || c-axis. Slightly above $T_c$, another QDW phase occurs and persists into the superconducting state shown in the pink area. Slightly down $T_c$, an AFM order with an odd-parity multipole has been revealed by NQR experiments labeled by a red star. i, The crystal structure of CeRh$_2$As$_2$ generated by Vesta [54]. ii, Schematic illustration of the bilayer Rashba-Hubbard model. Yellow circles represent the Ce atoms of CeRh$_2$As$_2$. The first- and second-neighbor intra-layer and inter-layer hopping integrals are $t$, $t'$ and $t_\perp$. iii, Physical properties of 4$f$ electron in Ce-based heavy-fermion systems. Large DOS near the Fermi level (left) and Schematic of the $c$-$f$ hybridization within the periodic Anderson model (right). iv Schematic drawing of the energy band nesting connected by a wave vector Q$_{nesting}$.

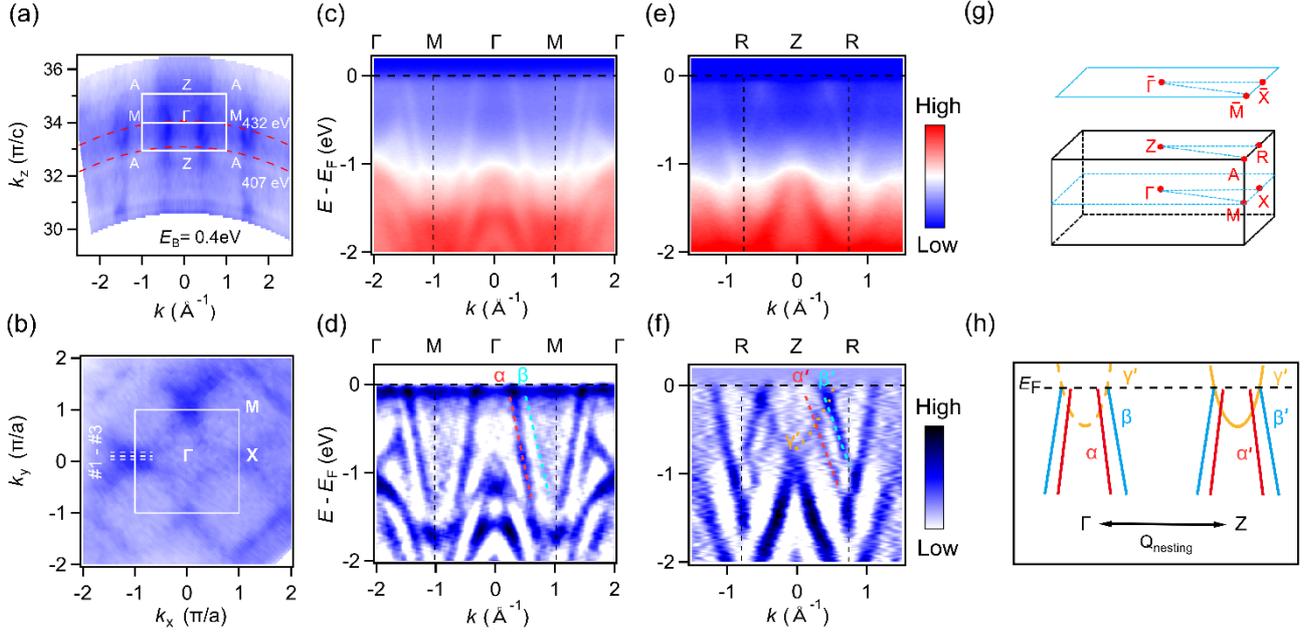

Fig. 2. Overall electronic structure of CeRh$_2$As$_2$. (a) Constant energy map of the ARPES data collected in the vertical Γ-M-Z-A plane with $E_B$ = 0.4 eV in a range of photon energies from 350 to 510 eV. The inner potential $V_0$=16.5 eV and the lattice constant $c$ = 9.8616 Å. (b) Fermi surface in the Γ-M-X plane. (c),(d) ARPES and curvature intensity map along high-symmetry line Γ-M respectively. (e),(f) ARPES and curvature intensity map along Z-R. Two quasi-2D hole pockets labeled by α and β and one strongly $k_z$-dispersive electron pocket γ′ surrounding Γ/Z points near the $E_F$ in (d),(f). (g) Three-dimensional bulk Brillouin zone (BZ) and the projected (001) surface BZ with high-symmetry points indicated. (h) Schematic illustration of the energy band nesting in CeRh$_2$As$_2$. All ARPES data in this figure are collected at 20 K.

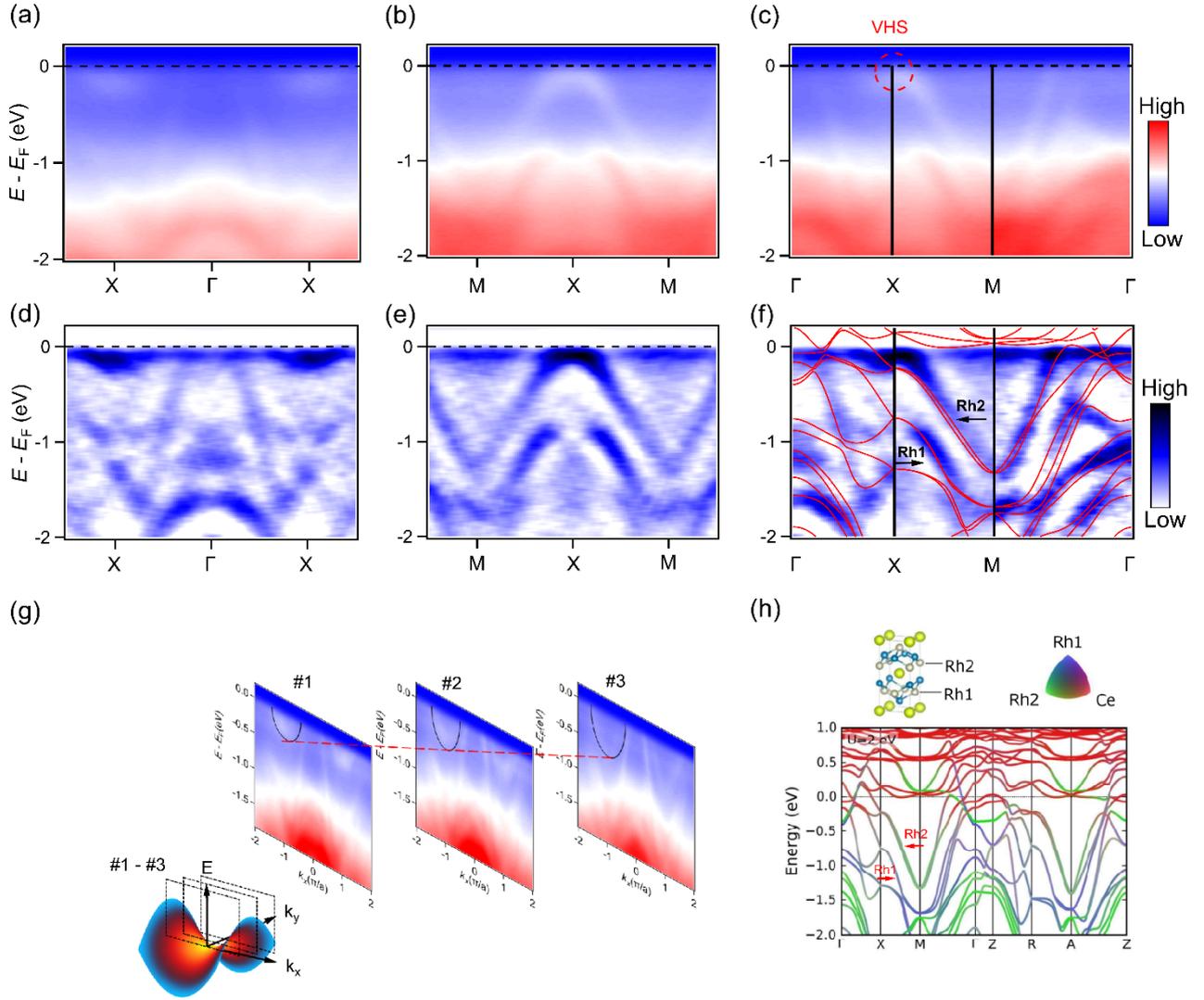

Fig. 3. VHS at the X point. (a)-(c) High-precision photoemission intensity plots along the Γ-X, X-M and Γ-X-M-Γ directions, respectively. (d)-(f) The corresponding curvature intensity plots. For comparison, the calculated band structure along Γ-X-M-Γ is superposed on the experimental data in (f). (g) A schematic diagram of VHS and ARPES intensity plots along the cuts #1 to #3, whose momentum locations are indicated by the white dashed lines in Fig. 2(b). (h) Band dispersion obtained by orbital- and atom-resolved DFT+U calculations for U = 2 eV. All ARPES data in this figure are collected at 20 K

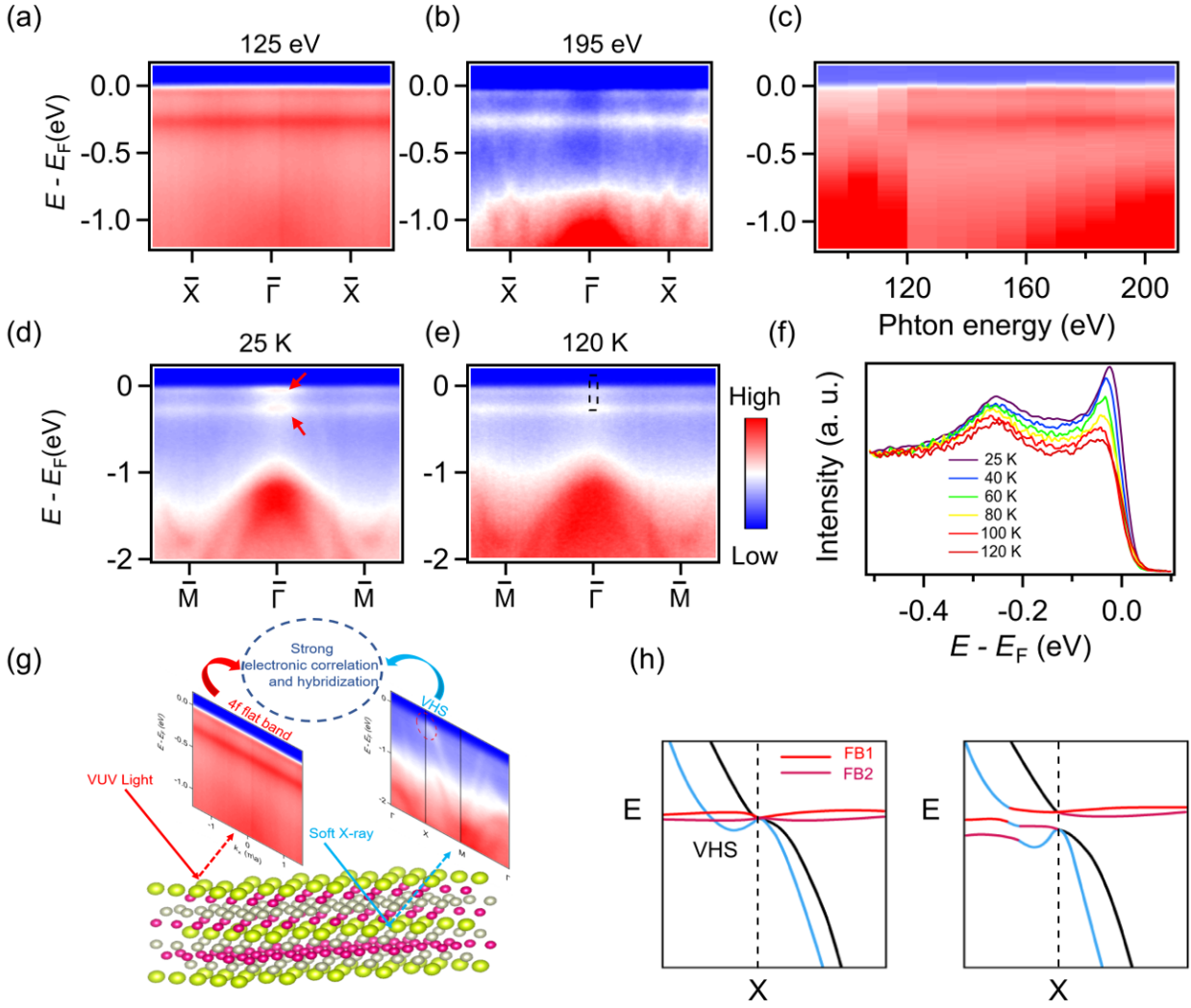

Fig. 4. Heavy 4*f* electron flat bands by wide range photon-energy and temperature dependent ARPES experiments. (a), (b) ARPES intensity plots of CeRh$_2$As$_2$ along $\bar{\Gamma}$-$\bar{X}$ at the labeled photon energy measured at temperature of 20 K. (c) ARPES data of CeRh$_2$As$_2$ along $\bar{\Gamma}$-$\bar{Z}$ at 20 K by photon energy-dependent ARPES measurements. (d), (e) ARPES data of CeRh$_2$As$_2$ along $\bar{\Gamma}$-$\bar{M}$ at 25 K and 120 K respectively, $h\nu$ = 175 eV with *p* polarization. (f) Temperature dependence of the EDCs in the vicinity of Γ labeled by black dashed rectangle in (e). (g) Summary of the interaction between 4*f* flat band and VHS. (h) Schematic band structure for a possible *f*-VHS (left) unhybridized and (right) hybridized case in CeRh$_2$As$_2$. Red lines illustrate nearly flat *f*-electron bands, and blue and black lines show conduction Dirac bands with VHS. There should be two bands for both conduction electrons and f-electrons because of the sublattice degree of freedom in locally noncentrosymmetric systems. Band degeneracy at the X point is protected by nonsymmorphic space group symmetry.